# Development of Surrogate Methods for Energy Production Process Water Characterization

**Babajide Kolade** [1*]

[1] Consultant; jide.kolade@gmail.com
* Correspondence: jide.kolade@gmail.com

**Abstract:** Effluent streams of process water used in energy production are contaminated with organic compounds which limits reusability of the water streams. Energy producers develop expensive monitoring and treatment methods to limit impact of the contamination on production. Standard methods for quantifying dissolved organics is process-affected water is laborious and time-consuming. Results from detailed characterization of process water with high concentration of dissolved organics are presented. Methods applied to characterize effluent stream include gravimetric analysis, elemental analysis, and LCMS. The results are used to develop efficient surrogate methods that may be used in continuous improvement in energy production operations.

## 1. Introduction

Effluent streams of process water used in energy production are contaminated with inorganic and organic compounds such as naphthenic acids. The contamination limits reusability and recyclability of the water streams in the production process and imposes constraints on disposal due to its significant adverse environmental effects. Energy producers develop extensive water monitoring and treatment methods to ameliorate production and environmental impacts. Treatment of process-impacted water streams depends on high-quality characterization of the water streams and quantification of components. Methods used for quantifying dissolved organics in process water streams include: Fourier transform infrared spectroscopy (FTIR) [1,2], gas chromatography-mass spectrometry (GCMS) [3,4], high performance liquid chromatography electrospray ionization mass spectrometry (HPLC-ESI-MS) [5,6], and liquid chromatography time-of-flight mass spectrometry (LC-TOF-MS) [7–9]. Comparisons of the different analytical methods [2,5,10] show that high-resolution methods yield lower standard error and are more selective; that calibration standards influenced quantitation; and that different sample preparation methods did not statistically affect total concentration.

Standard methods of characterizing naphthenic acid fractions in process-affected water, performed by a qualified analyst, is laborious and time-consuming.   The objective of this study is to develop an efficient surrogate method to quantify key constituents of process-affected water that may be applied within a continuous monitoring and improvement operational framework.

## 2. Materials and Methods

*Sample Collection*

Steam generator blowdown was collected from the client's oil sands operation in January 2012 and December 2013. Blowdown is the effluent stream of the steam generation process and contains concentrated amounts of dissolved organics. The steam plant water recycling rate is a controlling factor in the total amount and type of dissolved organics in the blowdown. The collected process-affected streams were stored in 55-gallon drums under nitrogen blanket till they reached the laboratory. At the laboratory, the drums were turned in a mechanical drum roller to homogenize the contents. Samples of collected effluent streams for the different years were partitioned and stored in amber bottles under nitrogen blankets for subsequent analyses. These samples are referred to as 2012 and 2013 blowdown samples in the remainder of the text. The 2012 and 2013 samples had nominal as-is pH of 11.7 and 11.9, respectively.

*Gravimetric Analysis*

2012 and 2013 blowdown samples were neutralized using hydrochloric acid (HCl) solutions of varying concentrations to specified pH levels between pH 11 and 2. Volumes of HCl used in neutralization were tracked and resulting dilution effects were accounted for in subsequent analyses and in reported in the results. Neutralized samples were held in closed sample bottles for soak times of two hours and filtered through 47mm Millipore 0.45-micron hydrophilic PTFE filters (FHLC04700) using a vacuum filtration system that was under an initial vacuum of 10-in. of Hg.

Starting blowdown sample sizes varied between 10 grams and 100 grams for the different pH levels. The goal was to precipitate a maximum of 20 milligrams of solids on the filters to prevent clogging of filters but get enough precipitate solids that would minimize experimental uncertainty. Dry filters that were desiccated for at least 24 hours were used in these experiments. The filters were not prewashed because it was confirmed that sodium hydroxide (NaOH) and HCl solutions between pH 12 and pH 2 do not react with or dissolve these filters to alter their weights, and the prewashing step may induce additional experimental uncertainty. Filters were not rinsed post-filtration because rinsing them alters solubility of the solids and reduce precipitate amounts. Then, filters and precipitates were dried in an oven at 103°C for one hour and precipitate weights were measured. Five repeat experiments at each pH level were conducted. The modified Thompson-Tau outlier detection method described in ASME's test uncertainty code [11] was used to identify spurious data points from repeat measurements. Statistical measures and uncertainty estimates were calculated from remaining data points.

Filtrates from repeat experiments were combined for each pH level and stored in amber sample bottles under nitrogen blankets. Density, conductivity, Total Carbon (TC), Total Organic Carbon (TOC), elemental composition, and Ultraviolet-Visible (UV-Vis) spectrophotometric absorbance values of filtrate samples were measured. In order to get consistent TOC measurements, filtrates analyzed for TOC were neutralized to pH 12 and repeat measurements reported by the analyst were checked for spurious data points. Elemental compositions were determined by Inductively Coupled Plasma Atomic Emission Spectroscopy (ICP-AES) and Ion Chromatography (IC) methods. Additionally, naphthenic acids concentrations and spectra of filtrate samples were analyzed using a Liquid Chromatography-Mass Spectroscopy (LCMS) method.

*LCMS Method*

LCMS was used to identify and quantify classes of naphthenic acids in the samples. Samples were spiked with an isotopically labeled internal standard and adjusted to pH 11. Samples were injected into a high-performance liquid chromatography column where organic components were separated and eluted with a methanol-water gradient program using a nonpolar stationary phase. Eluted components were ionized using electrospray ionization (ESI) and detected in a mass spectrometer based on measurements of their mass-to-charge (m/z) values. Naphthenic acids signal magnitudes (abundance) across the entire m/z scale were corrected by applying experimentally determined response factors. Response factors account for differences in ionization efficiencies of different chain lengths. The m/z values were used for identification of constituent naphthenic acids and ratios of their corrected abundance to the internal standard were used to determine their concentrations. Background and details of the method are documented in these references [12–14].

*UV-Vis Spectroscopy*

UV-Vis absorbance spectroscopy for process water measures relative absorbance of light as a function of wavelength as it interacts with the dissolved compounds in the water. Peak absorption wavelength for each dissolved compound depends on its atomic and molecular structure. Lower molecular weight phenolic compounds and naphthenic acids have peak absorbance in the UV range (200 nm – 350 nm), while larger molecular weight phenolic compounds and other organics have peak absorbance in the visible range (400nm – 770nm) [15,16]. Absorbance peaks of different dissolved compounds overlap and form a continuous spectrum for a process water sample. Differences in spectra and derivatives of spectra yield qualitative and quantitative information about classes of compounds dissolved in one sample as compared to another sample.

UV-Vis absorbance measurements were carried out on an Agilent Technologies Cary 100 UV-Vis Spectrophotometer using the following scanning parameters: a scan rate of 600 nm/min, at a data interval of 1.00 nm, with an average time of 0.10 millisecond. Filtrate samples were diluted 10x and 120x using Millipore Milli-Q filtered water adjusted to pH 12. Scans were acquired for the diluted filtrate samples using filtered water as background between the 200 nm and 800 nm interval.

*Uncertainty Analysis*

Prior to starting the experiments, measurement processes were characterized to determine distribution of aleatoric uncertainty associated with each process. These distributions were used to calculate uncertainty estimates of mean values. But more importantly, they were used to alert the analyst of a potential systematic error and avoid wasted experiments. This characterization process was used to avoid cascading systematic errors in experimental results.

For example if a weighing process is integral to the experimental data quality, an analyst may repeat the process of weighing a representative sample (e.g., filter paper, fouling coupon) enough times to generate a distribution and calculate a valid standard deviation of the distribution. Using the standard deviation and the Student's t-table at a specified confidence interval, one calculates the maximum difference between two consecutive values from the weighing process that may be attributed to randomness. The analyst now needs only to weigh samples twice during the experiments. If the difference between these consecutive weights exceeds the predetermined maximum, the analyst is alerted of a potential systematic error and would fix the error before continuing the experiments.

Individual uncertainty sources in the gravimetric analysis such as accuracy and readability of the weighing balance, homogeneity of sample material, and effects of the drying method and lab humidity were identified and evaluated. These uncertainty values were combined using the Root-Sum-Square (RSS) technique at a 95% confidence interval and reflected in error bars in plots and in tabulated results.

## 3. RESULTS & DISCUSSION

Amount of precipitates in 2012 samples were higher than the amount of precipitates in 2013 samples because of higher blowdown recycle rate in 2012. As shown in Figure 1, coloration of filtrate samples decreased with decreasing pH levels from dark and opaque through amber to yellow. Correspondingly, amounts of precipitates increased with decreasing pH levels. The gravimetric plot in Figure 2 shows that between pH 10 and pH 7.5, 2012 and 2013 blowdown samples precipitated 30 ppm and 20 ppm of solids, respectively. Figure 2 also shows that amounts of filterable precipitates jump around pH levels 10, 7, and 6. These pH levels appear to be transition points in solubilities of different components of the steam generator blowdown. Amounts of precipitates increase asymptotically till the lowest pH of 2.

Table 1 summarizes results of the characterization of blowdown sample and gravimetric analysis. "*pH 2 TOC*" is the concentration of organic carbon present in the sample filtrate neutralized to a pH level of 2. Proportion of precipitates at pH 2 (termed "*pH 2 Insolubles*") is an indication of amounts of larger molecular weight organics originally soluble in the sample. Amounts of pH 2 insoluble fractions were found to be 3,000 ppm for 2012 blowdown and 1,400 ppm for 2013 blowdown.

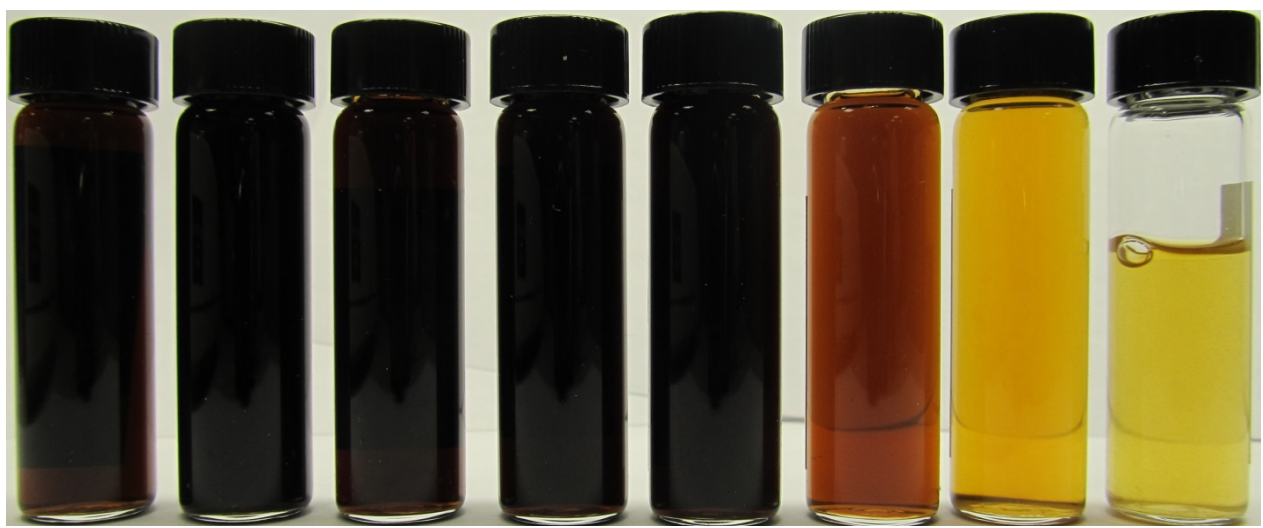

Figure 1. 2013 Blowdown filtrate samples: as-is, pH 11.8, 10, 8.9, 7, 6, 4, 2

Table 1. Characterization of As-received Blowdown Samples

| Sample Name | 2012 Sample | 2013 Sample |
|---|---|---|
| Density (g/cm$^3$) | 1.01454 ± 1E-5 | 1.01040 ± 1E-5 |
| pH | 11.66 ± 0.06 | 11.86 ± 0.06 |
| TOC (mg/L) | 4,600 ± 200 | 2,290 ± 50 |
| TC (mg/L) | 4,700 ± 200 | 2,390 ± 50 |
| pH 2 Insolubles (mg/L) | 2,970 ± 50 | 1,390 ± 40 |
| pH 2 TOC/ TOC | 49% | 66% |
| TDS (mg/L) | 27,900 | 20,300 |
| Conductivity (mS/cm) | 29 ± 2 | 23.3 ± 0.6 |

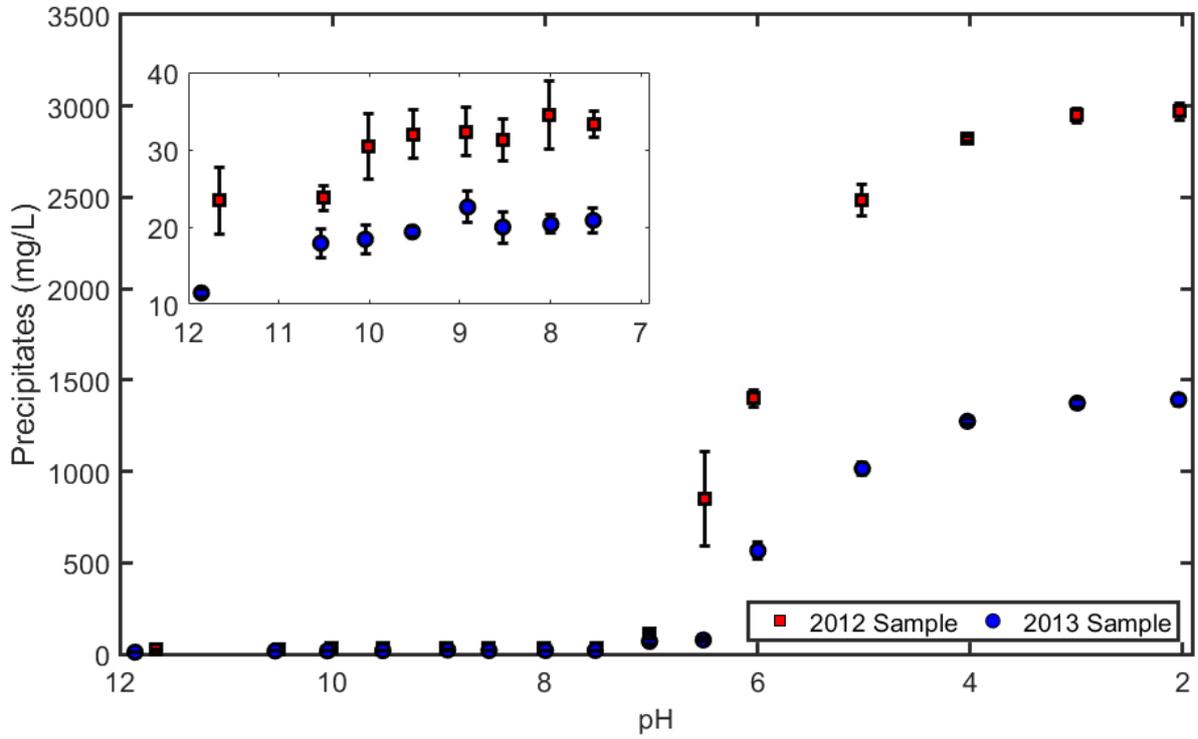

Figure 2. Concentrations of precipitates from pH neutralization of 2012 and 2013 blowdown samples; pH 7 neutralization yields 120 and 70 mg/L, respectively

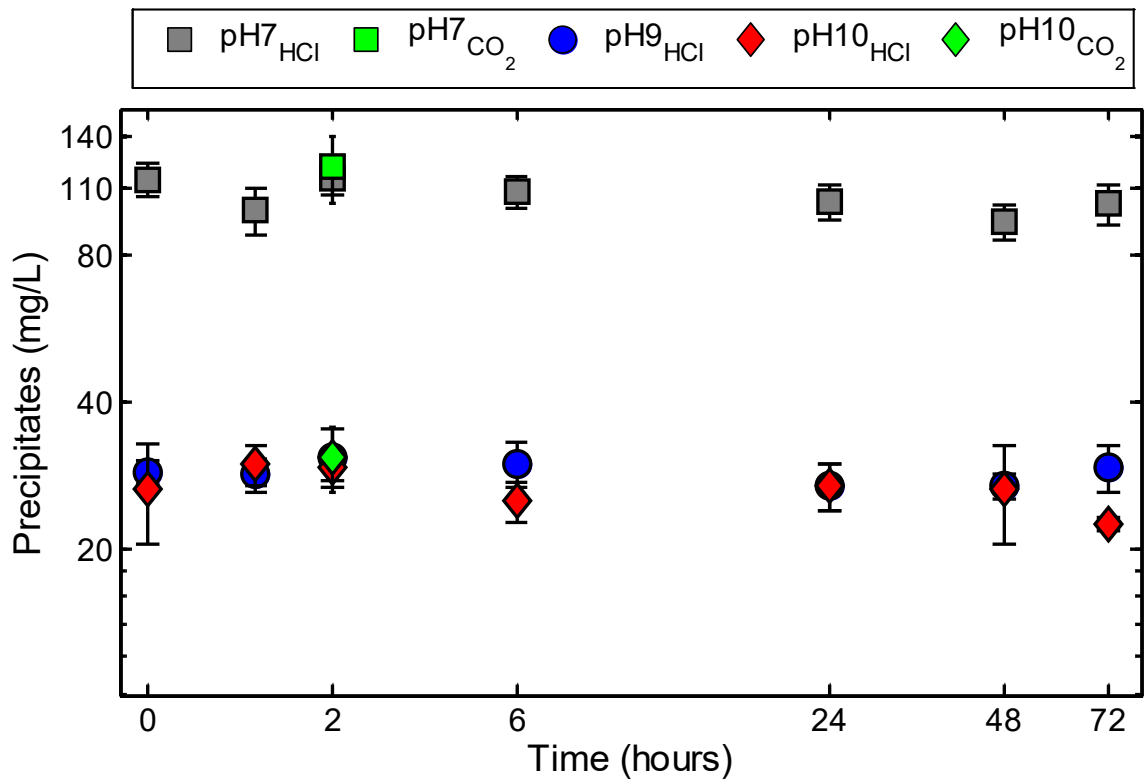

Figure 3. Effect of time and acid type on concentration of precipitates for 2012 blowdown samples

Soak time did not have a significant effect on the mass of precipitates derived from the gravimetric analysis. Gravimetric analyses of 2012 blowdown samples were conducted based on the experimental method described above at pH levels 10, 8.9, and 7, while using soak times of 0, 1, 2, 6, 24, 48, and 72 hours. As shown in Figure 3, amounts of precipitates for different soak times were all within experimental uncertainty. This result should be expected because protonation reactions which affect solubilities of organics in water are fast reactions, especially protonation by a strong acid like HCl. Figure 3 also shows that using $CO_2$ as the neutralizing acid, instead of HCl solutions, did not significantly affect the mass of precipitates derived from the gravimetric analyses. These results imply that operational processes devised to treat blowdown by acid clarification will not be kinetically limited.

*LCMS Results*

The chromatogram in Figure 4 from the LCMS analysis of 2012 blowdown shows elution times of different components in the sample. Naphthenic acids and small molecular weight (MW) organics elute first, within the 0 – 3 minutes time range, because of a combination of their polarity, MW and hydrophobicity. However, only naphthenic acids are detected in this time range because their ionization efficiencies are much higher. Components eluted in the naphthenic time range are considered in the mass spectral analysis and quantitation.

Figure 5 shows mass spectra of naphthenic acids fractions of filtrates from gravimetric analysis of 2012 blowdown. Centroids of the spectra are the Number-Averaged Molecular Weights (NAMW) of the filtrates. Number-averaged molecular weights of filtrates decrease with decreasing pH levels because larger MW naphthenic acids are filtered off. Figure 6 shows profiles of NAMW of naphthenic acids fractions of 2012 and 2013 blowdown filtrates as a function of pH. NAMW profiles for both years are similar indicating that the blowdown samples contain similar types of naphthenic acids. Naphthenic acids in blowdown samples have a maximum NAMW of 295 between pH 7 and pH 11.8 and a minimum NAMW of 230 at pH 2; NAMW profiles transition around pH 6.

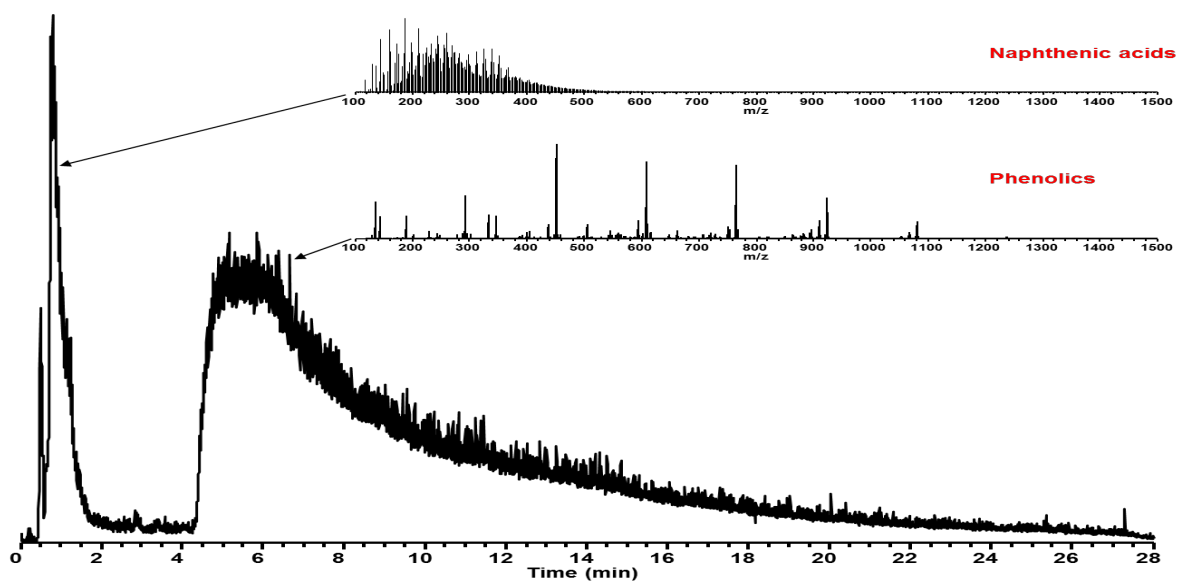

Figure 4. Chromatogram from LCMS analysis of 2012 Blowdown sample

7 of 15

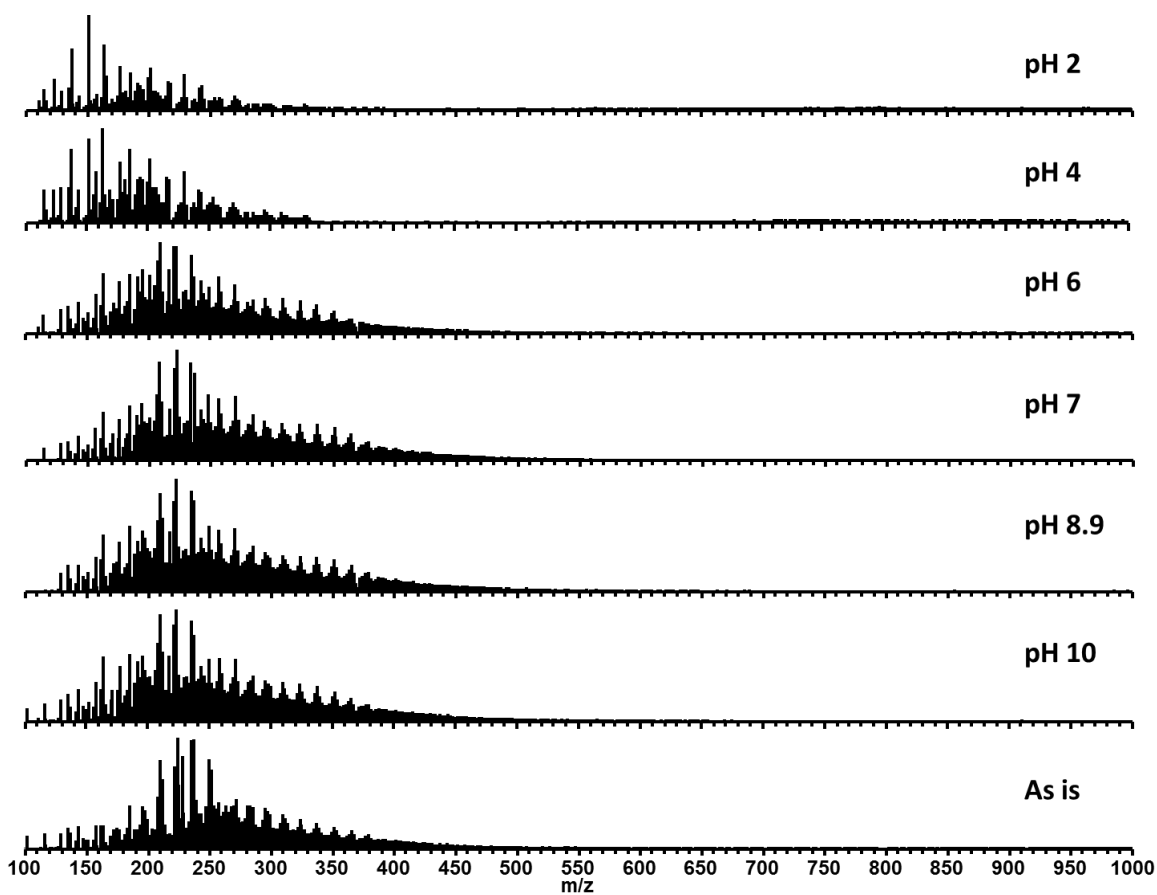

Figure 5. Mass spectra of naphthenic acids fractions of filtrates from gravimetric analysis of 2012 blowdown samples

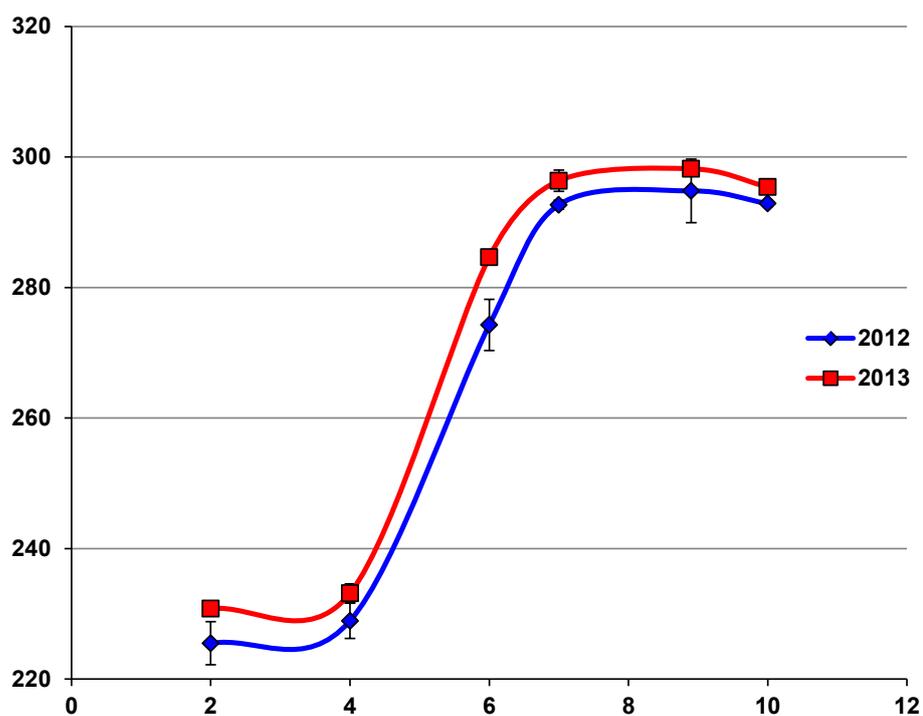

Figure 6. Number-averaged molecular weights of naphthenic acid fractions of filtrates from gravimetric analysis of 2012 and 2013 blowdown samples

Table 2 shows metrics calculated from LCMS analysis that compare naphthenic acids fractions of unfiltered 2012 and 2013 blowdown samples. Number-averaged molecular weight, average molecular weight, and typical structure of naphthenic acids in both years are similar. The average naphthenic acid had 23 carbon atoms and three condensed rings, and the most abundant naphthenic acid had 15 carbon atoms and two condensed rings. From Figure 6 and Table 2, one may conclude that 2012 and 2013 blowdown samples had the same types of naphthenic acids but 2012 blowdown had a larger fraction of phenolics or larger MW organics contributing to its TOC (See last row in Table 2).

Table 2. Naphthenic Acid Metrics Calculated from LCMS Analysis of Blowdown Samples

| Sample Name | Jan-2012 Blowdown | Dec-2013 Blowdown |
|---|---|---|
| LCMS Naphthenic Acid Analysis - $C_nH_{2n+z}O_2$ | | |
| Number-Averaged Molecular Weight | 293 | 294 |
| Average Molecular Weight | 351 | 354 |
| H/C Molar Ratio | 1.74 | 1.74 |
| Average (n, z) | (23.1, -5.8) | (23.4, -5.8) |
| Naphthenic Acid Carbon/TOC | 0.52 | 0.61 |

*UV-Vis Results*

Figure 7 shows the UV-Vis absorbance spectra of 2012 blowdown filtrates. UV-Vis absorbance spectra are not particularly remarkable by themselves. As expected, absorbance spectra decrease with decreasing pH because material is filtered off. Also, 2012 blowdown samples have higher absorbance spectra than 2013 blowdown samples because of higher TOC content.

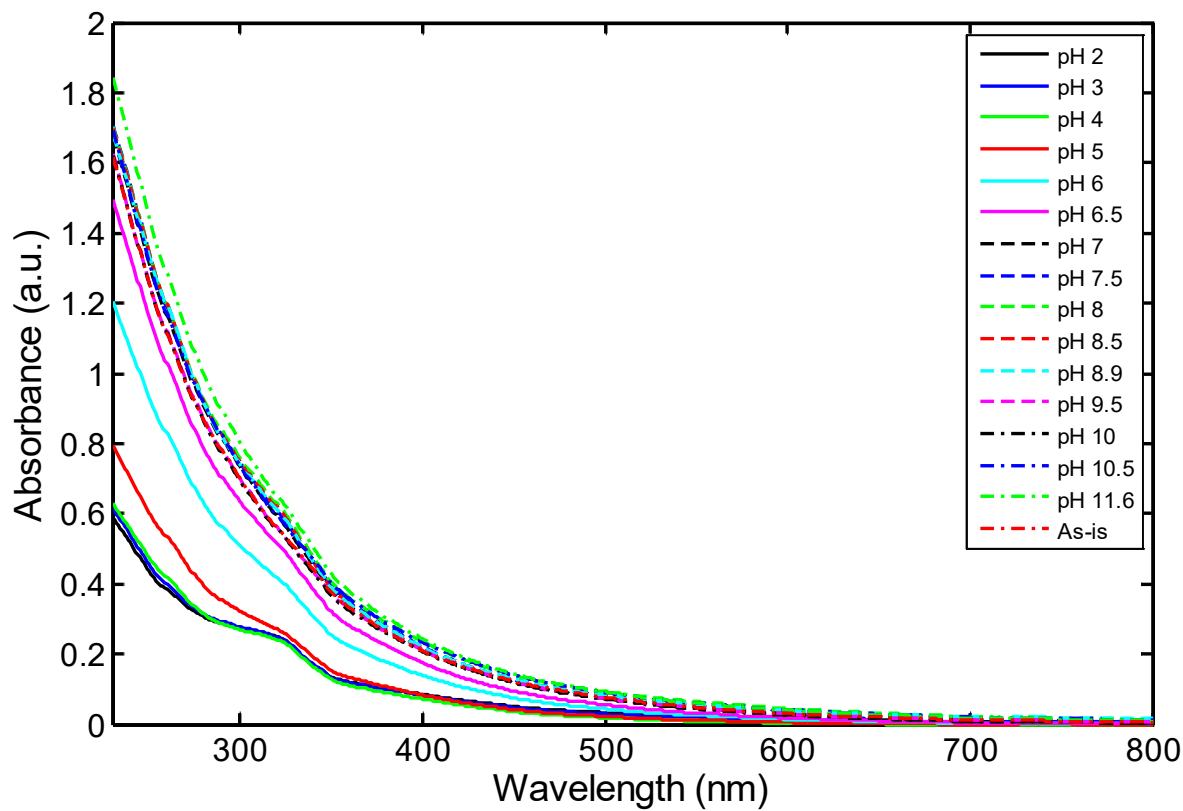

Figure 7. UV-Vis absorbance spectra of 2012 blowdown filtrates

## 4. Discussion

Measurements of TOC and elemental compositions for unfiltered samples and filtrates at different pH levels were used to determine compositions of precipitates. This approach of comparing what is in the water before and after filtration to determine precipitate composition was deemed more accurate than extracting and analyzing precipitates from filter papers because the latter is a multistep process that incurs additional uncertainties. Inorganic species were a major portion of precipitates between pH levels 10.5 and 7.5, whereas precipitates were almost completely organic at pH levels lower than 7. Though total amount of precipitates was low in the higher pH range as shown in Figure 5, inorganics account for 40 – 90% of precipitates within the range. This may be because there is a significant change in solubilities of divalent and multivalent cations, such as Ca, Mg, and Si, in blowdown as pH decreases. For example, the concentration of calcium ion in 2012 blowdown is 4.6 ppm at pH 10.5 and 1 ppm at pH 7. In contrast, monovalent ions such as sodium and potassium ions appear to be insensitive to pH levels. Figure 10 and Figure 11 show the differences in solubilities of the cations. The striking change in solubilities of the multivalent cations and differences in the compositions of precipitates may have significant impact on blowdown disposal because a reduction in pH of the disposal stream may clog filter cartridges or silt out injection wells.

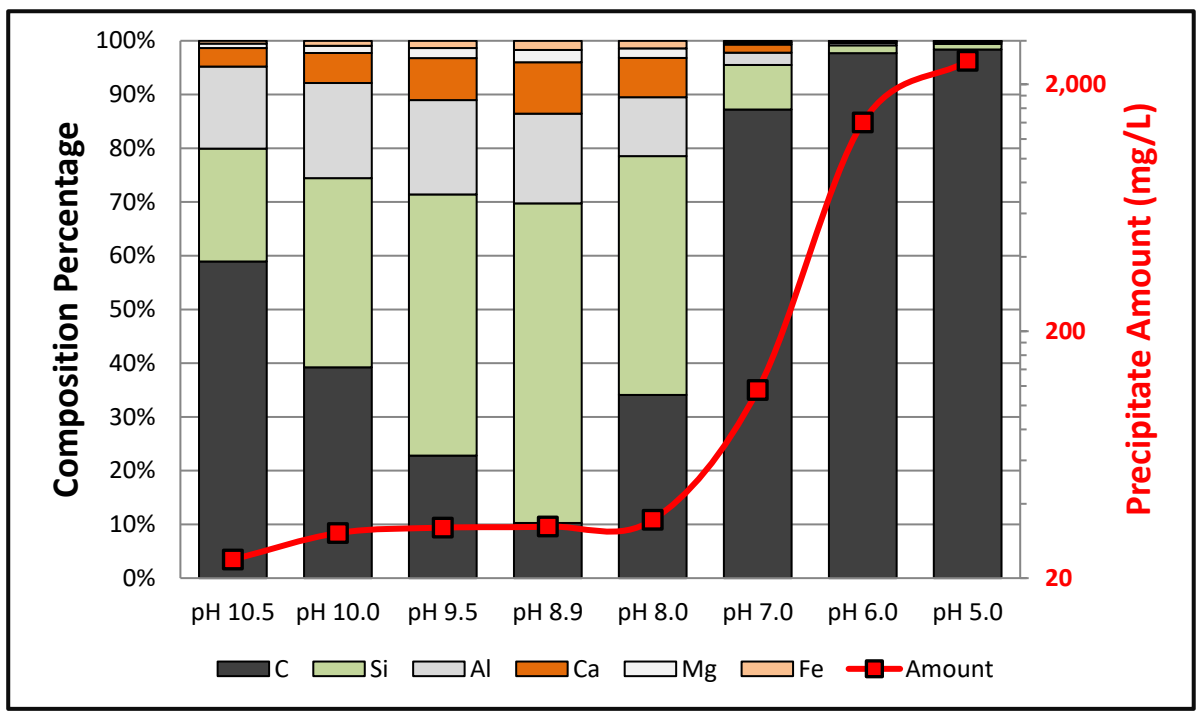

Figure 8. Composition of precipitates from pH neutralization of 2012 blowdown sample, on an oxygen-free, sulfur-free basis; precipitate amount on right axis

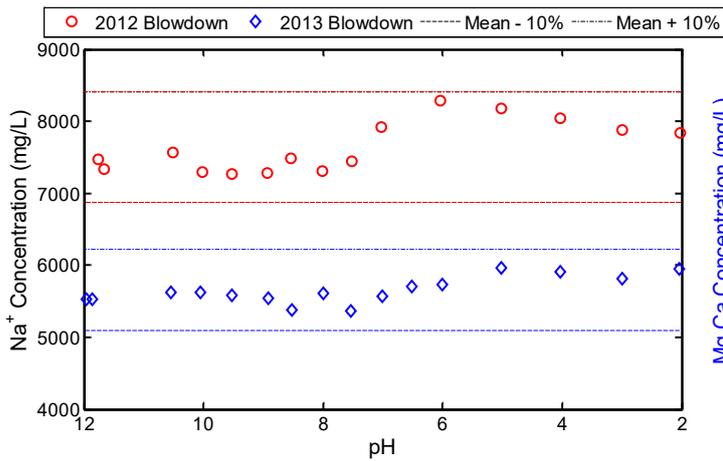

Figure 9. Concentrations of sodium ion in filtrates

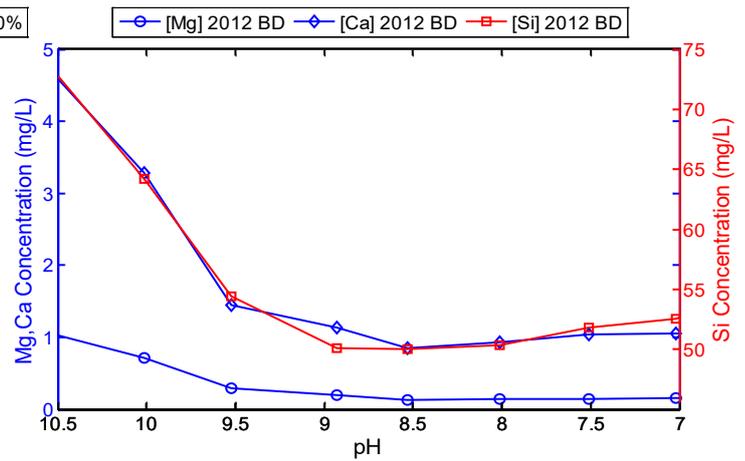

Figure 10. Concentrations of multivalent ions in filtrates

Mass balance equations and least-squares regression may be used to estimate the average molecular oxygen to carbon (O/C) ratio of the dissolved organics. The following assumptions are needed to complete the analysis: First, that precipitates from the gravimetric analysis are mainly organics. Figure 8 shows that this assumption is largely valid at a pH level 7 and lower. Second, that dissolved organics are comprised only of carbon, hydrogen, and oxygen. This assumption implies that the blowdown stream has negligible organosulfur and organic nitrates fractions, or that the mass fraction of sulfur and nitrogen in these compounds are minimal. Third, that the molecular hydrogen to carbon (H/C) ratio is a constant value. A constant value of 1.33 was used in the analysis, which is based on an average of typical H/C ratios in naphthenic acids and phenols. Note that an error of magnitude 1 in the assumed H/C ratio only yields an error of magnitude 0.06 in the estimated O/C ratio because of the difference in atomic weights. Given these assumptions, one may write mass balance equations at each pH level relating the total amount of organics to measured precipitate weights, measured TOC, assumed hydrogen fraction, and estimated oxygen fraction; these equations are shown below.

Total amount of organics in a sample is equated to the measured weight of precipitates, total organic carbon in its filtrate, hydrogen fraction in the filtrate, and oxygen fraction in the filtrate at each pH level. The mass balance equations shown below are used to estimate the average molecular O/C ratios and total concentration of organics in 2012 and 2013 blowdown samples.

$$Total\ Organics\ in\ Unfiltered\ Sample = Precipitates + Filtrate\ C + Filtrate\ H + Filtrate\ O$$

$$Total\ Organics = Precipitates_{@pH\ 2} + TOC_{@pH\ 2} + \overbrace{\frac{4}{3}*\frac{1}{12}*TOC_{@pH\ 2}}^{Hydrogen\ fraction\ in\ filtrate} + \overbrace{O/C*\frac{16}{12}*TOC_{@pH\ 2}}^{Oxygen\ fraction\ in\ filtrate}$$

$$Total\ Organics = Precipitates_{@pH\ 3} + TOC_{@pH\ 3} + \frac{4}{3}*\frac{1}{12}*TOC_{@pH\ 3} + O/C*\frac{16}{12}*TOC_{@pH\ 3}$$

$$\vdots$$

$$Total\ Organics = Precipitates_{@pH\ 10} + TOC_{@pH\ 10} + \frac{4}{3}*\frac{1}{12}*TOC_{@pH\ 10} + O/C*\frac{16}{12}*TOC_{@pH\ 10}$$

Rearranging these equations in matrix-vector form yields:

$$\begin{bmatrix} 1 & -4/3 * TOC_{@pH\,2} \\ 1 & -4/3 * TOC_{@pH\,3} \\ 1 & \vdots \\ 1 & -4/3 * TOC_{@pH\,10} \\ 1 & \vdots \end{bmatrix} \begin{bmatrix} Total\ Organics \\ O/C \end{bmatrix} = \begin{bmatrix} Precipitates_{@pH\,2} + TOC_{@pH\,2} + 1/9 * TOC_{@pH\,2} \\ Precipitates_{@pH\,3} + TOC_{@pH\,3} + 1/9 * TOC_{@pH\,3} \\ \vdots \\ Precipitates_{@pH\,10} + TOC_{@pH\,10} + 1/9 * TOC_{@pH\,10} \\ \vdots \end{bmatrix}$$

Least-squares regression was used to solve the matrix-vector equation above. The average molecular O/C ratios were 0.27 and 0.45 for 2012 and 2013 blowdown samples and total concentrations of organics were 6,400 ppm and 3,800 ppm, respectively. Figure 11 shows a breakdown of the dissolved organics in 2012 blowdown based on this analysis. Organics in 2013 blowdown have a higher O/C ratio than 2012 blowdown and, hence, are more soluble [17]. This conclusion is corroborated by experimental data (see Table 1) that shows that the 2013 sample had a higher percentage of its TOC still soluble (66%) in the pH 2 filtrate compared to 2012 blowdown sample (49%). This percentage is a useful and easy-to-quantify metric for tracking water quality and to design operational plans to mitigate against organic boiler fouling issues.

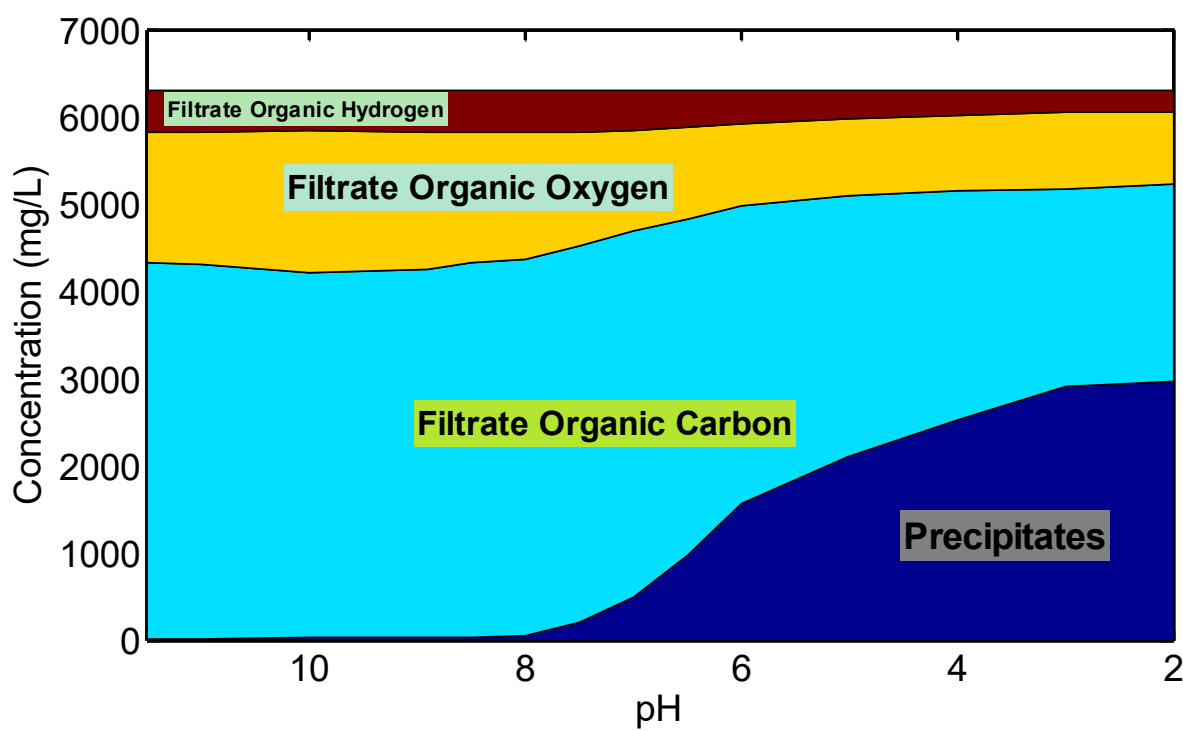

Figure 11. Breakdown of dissolved organics in 2012 blowdown sample based on mass balance

*UV-Vis Discussion*

Absorbance spectra for all filtrates for both the 10x and 120x dilutions were partitioned into four regions (200 – 250 nm, 250 – 350 nm, 350 – 400 nm, and 400 – 800 nm) and the average absorbance within each region was calculated. Average absorbance values in the 200 – 250 nm range and 250 – 350 nm for 10x dilutions were discarded because signal values were too high, and the spectrophotometer was saturated. Partitioning the UV-Vis spectra highlights different parts of the spectra that may correlate to other quantifiable characteristics of the dissolved organics in blowdown samples.

As shown in Figure 12 and Figure 13 respectively, average absorbance values in the 200 – 250 nm range for 120x dilution correlated remarkably well with NAMW of naphthenic acid fractions measured by LCMS and average absorbance values in the 350 – 400 nm range for 10x dilution correlated well with TOC. These correlations were exploited to develop statistical predictors of TOC, naphthenic acids NAMW, and naphthenic acids carbon concentration that works for both years over the entire range of pH values.

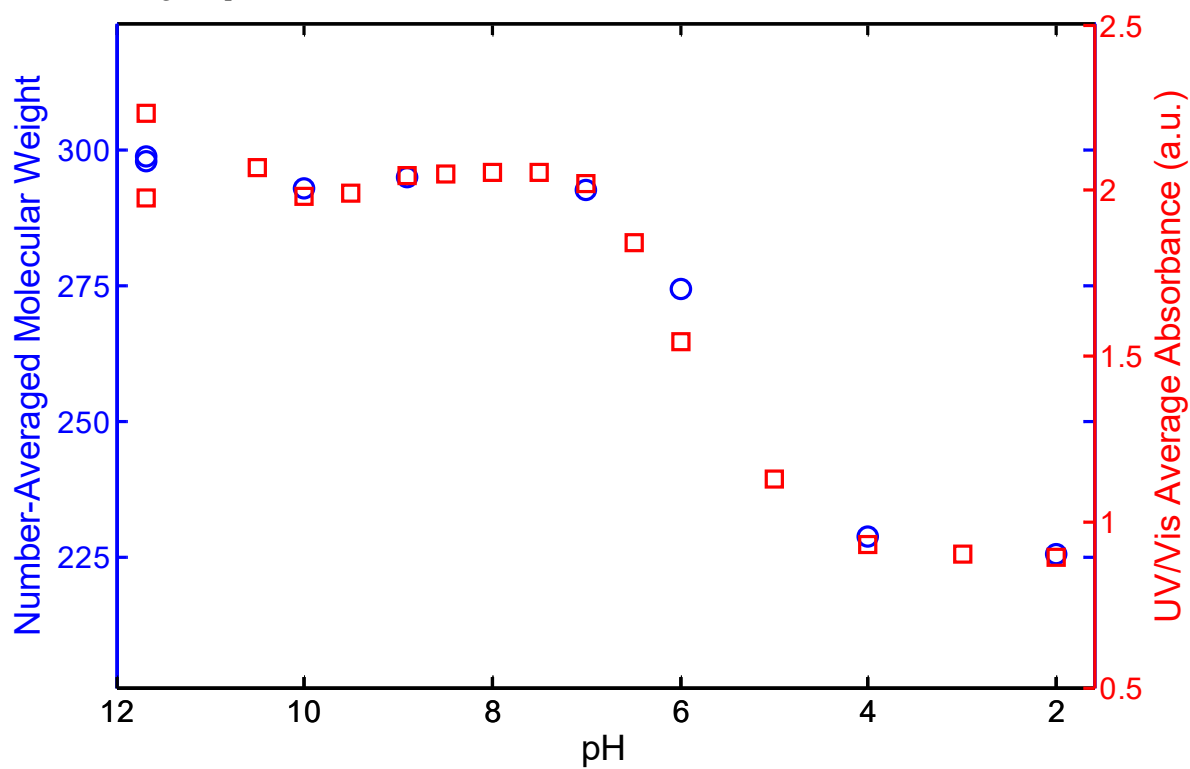

Figure 12 – NAMW of naphthenic acids fractions and average UV-Vis absorbance in the 200-250 nm range for 120x dilution of filtrates of 2012 blowdown

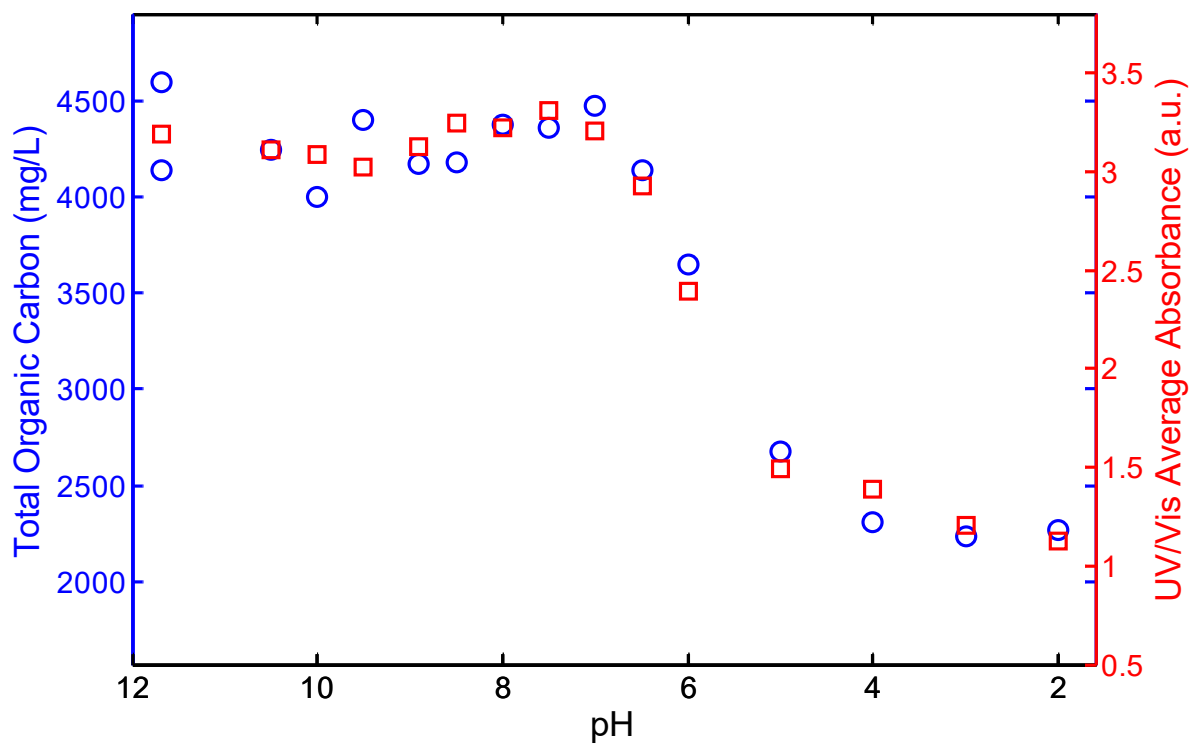

Figure 13. TOC and average UV-Vis absorbance in the 350-400 nm range for 10x dilution of filtrates of 2012 blowdown

Equations 1 through 3 are statistical models for TOC, NAMW and carbon fraction of naphthenic acids in blowdown samples. Goodness-of-fit metrics shown in Table 3 indicate that these equations are good predictors of some characteristics of the organics in blowdown samples. These equations may not perform well in predicting the organic content for other process water because of the limited data set on which they were developed. This concept of correlating integral quantities from relatively cheap and accessible UV-Vis measurements to more complex characterizations of organics in process water may be applied to other quantities of interest and for other classes of process-affected water. A stand-alone numerical script that imports and analyzes output data from a Cary UV-Vis Spectrophotometer may be developed to facilitate development of other correlations. Correlations presented here and yet-to-be-developed correlations may be used as continuous improvements metrics that can be tracked by energy producers to optimize operations.

$$\frac{TOC}{4000\ ppm} = 0.33 + 0.44 x_1 - 2.68 x_2 + 0.55 x_1 x_2 \tag{1}$$

$$\frac{NAMW}{300} = 0.64 + 0.39 x_3 + 1.70 x_4 - 1.04 x_3 x_4 + (0.59 x_3 - 1.32)\frac{TOC}{4000} \tag{2}$$

$$\frac{Naph\ Acid\ C}{4000\ ppm} = 0.13 - 0.55 x_3 + 2.47 x_4 + (2.16 x_3 - 4.90 x_4 - 0.96)\frac{TOC}{4000} \tag{3}$$

Where:
$x_1$: Average UV-Vis absorbance in the 350 – 400 nm range for 10x dilution
$x_2$: Average UV-Vis absorbance in the 400 – 800 nm range for 10x dilution

$x_3$: Average UV-Vis absorbance in the 200 – 250 nm range for 120x dilution

$x_4$: Average UV-Vis absorbance in the 250 – 350 nm range for 120x dilution

Table 3. Goodness-of-fit Metrics for Statistical Models

|  | # Observations (DOF) | RMSE | $R^2$ | *p*-value |
|---|---|---|---|---|
| TOC Model (Eq. 1) | 32 (28) | 0.0715 | 0.939 | 3.8E-17 |
| NAMW Model (Eq. 2) | 16 (10) | 0.0211 | 0.968 | 3.7E-07 |
| Nap. Acid Carbon Model (Eq. 3) | 14 (8) | 0.0322 | 0.978 | 1.9E-06 |

**5. Conclusions**

The detailed characterization of energy production process-affected water and development of surrogate methods for monitoring water quality led to the following conclusion:

1. Precipitation time or type of neutralizing acid did not have a significant effect on amounts of precipitates. So conceptual processes that propose to treat steam generator blowdown through acid clarification should not be kinetically limited.
2. Inorganic species were a dominant portion of the precipitates at pH levels between 10.5 and 7.5, whereas precipitates were mainly organic at pH 7 and lower. This finding may have significant impact on blowdown disposal because a reduction in pH of the disposal stream may clog filter cartridges or silt out injection wells.
3. Organics in 2013 blowdown had a higher molecular oxygen-to-carbon ratio than organics in 2012 blowdown and therefore were more soluble. Correspondingly, 2013 blowdown had a higher fraction of its TOC still soluble in the pH 2 filtrate compared to 2012 blowdown.
4. Number-Averaged Molecular Weights (NAMW) of naphthenic acids in filtrates decreased with decreasing pH levels because larger MW naphthenic acids were filtered off. From mass spectra data, the main transition point for NAMW was around pH 6.
5. 2012 and 2013 blowdown samples had the same types of naphthenic acids, but 2012 blowdown had a larger fraction of phenolics or larger MW organics contributing to its TOC.
6. UV-Vis absorbance spectra correlated remarkably well with NAMW of naphthenic acids and TOC of filtrates. Statistical models developed from these correlations were good predictors of characteristics of the organics in blowdown across both years.

Supplementary Materials: Not applicable.

Funding: This research received no external funding.

Acknowledgements: The author would like to acknowledge Judith Waters and Levi Standeford for their assistance with sample analyses. The author is grateful to Leonard Nyadong for his assistance with sample analysis and insightful discussions.

Author Contributions: Not applicable.

Data Availability Statement: Not applicable.

Conflict of Interest: The author declares no conflict of interest.